\documentclass[amsmath,superscriptaddress,reprint,aps]{revtex4-2}
\usepackage{graphicx}% Include figure files
\usepackage{dcolumn}% Align table columns on decimal point
\usepackage{bm}% bold math
\usepackage{hyperref}% add hypertext capabilities
\usepackage{siunitx}
\usepackage{color, soul}

\setstcolor{red}
\usepackage[mathlines]{lineno}% Enable numbering of text and display math
\begin{document}
	
	%\preprint{APS/123-QED}
	
	\title{Versatile tuning of Kerr soliton microcombs in crystalline microresonators}% Force line breaks with \\
	%%Versatile tuning of Kerr soliton microcomb with spontaneous coupling enhancement in crystalline microresonators

	\author{Shun~Fujii}
	\email[Corresponding author. ]{shun.fujii@phys.keio.ac.jp}	
	\affiliation{Department of Physics, Faculty of Science and Technology, Keio University, Yokohama, 223-8522, Japan}
	\affiliation{Quantum Optoelectronics Research Team, RIKEN Center for Advanced Photonics, Saitama 351-0198, Japan}
	\affiliation{Department of Electronics and Electrical Engineering, Faculty of Science and Technology, Keio University, Yokohama, 223-8522, Japan}

	\author{Koshiro~Wada}
	\affiliation{Department of Electronics and Electrical Engineering, Faculty of Science and Technology, Keio University, Yokohama, 223-8522, Japan}

	\author{Ryo~Sugano}
	\affiliation{Department of Electronics and Electrical Engineering, Faculty of Science and Technology, Keio University, Yokohama, 223-8522, Japan}
	
	\author{Hajime~Kumazaki}
	\affiliation{Department of Electronics and Electrical Engineering, Faculty of Science and Technology, Keio University, Yokohama, 223-8522, Japan}
	
    \author{Soma~Kogure}
	\affiliation{Department of Electronics and Electrical Engineering, Faculty of Science and Technology, Keio University, Yokohama, 223-8522, Japan}
	
		\author{Yuichiro~K.~Kato}
	\affiliation{Quantum Optoelectronics Research Team, RIKEN Center for Advanced Photonics, Saitama 351-0198, Japan}
	\affiliation{Nanoscale Quantum Photonics Laboratory, RIKEN Cluster for Pioneering Research, Saitama 351-0198, Japan}

    \author{Takasumi~Tanabe}
    \email[Corresponding author. ]{takasumi@elec.keio.ac.jp}
	\affiliation{Department of Electronics and Electrical Engineering, Faculty of Science and Technology, Keio University, Yokohama, 223-8522, Japan}

%	\date{\today}% It is always \today, today,
	% but any date may be explicitly specified
	
	\begin{abstract} %150 words or fewer
Microresonator-based optical frequency combs emitted from high-quality-factor microresonators, also known as microcombs, have opened up new horizons to areas of optical frequency comb technology including frequency metrology, precision sensing, and optical communication. To extend the capability of microcombs for such applications, large and reliable tunability is of critical importance.
Here, we show  broad spectral tuning of Kerr soliton microcombs in a thermally controlled crystalline microresonator with pump-detuning stabilization. The fundamental elements composing frequency combs, namely the center frequency, repetition frequency, and carrier-envelope offset frequency, are spectrally tuned by up to $-48.8$~GHz, $-5.85$~MHz, and $-386$~MHz, respectively, leveraging thermal effects in ultrahigh-Q crystalline magnesium fluoride resonators. We further demonstrate a 3.4-fold enhancement of soliton comb power resulting from thermal expansion with a temperature change of only 28~K by employing quantitative analyses of the fiber-to-resonator coupling efficiency.
	\end{abstract}

	\maketitle
	
	%\tableofcontents

	\section*{Introduction}	
	High-repetition rate microresonator-based frequency combs offer powerful and compact optical frequency comb (OFC) sources that are of great importance to various applications including a phase-coherent link for the optical to microwave domain~\cite{Spencer2018,Lucas2020}, optical communication~\cite{Marin-Palomo2017,Fujii:22}, spectroscopy~\cite{Suh600,Lucas2018}, and low-noise photonic microwave generation~\cite{Liang2015:high,PhysRevLett.122.013902,Yang2021,Kwon2022}. In particular, the soliton mode-locked states of microresonator frequency combs provide high coherence and smooth spectral profiles~\cite{herr2014temporal}, thereby making them highly desirable for most applications~\cite{Spencer2018,Marin-Palomo2017,Suh600,Liang2015:high,Suh884,Trocha887,Lucas2018,Lucas2020,Fujii:22,PhysRevLett.122.013902,Yang2021}. One significant difference between the microcombs and conventional OFC sources is their resonator configuration~\cite{Fortier2019,doi:10.1126/science.aay3676}. Most mode-locked lasers are based on composite systems with a gain medium, bulky free-space optics, active and passive optical fibers~\cite{doi:10.1126/science.aay3676}; moreover, laser-cavity actuators such as temperature control systems, piezoelectric transducer devices, or high bandwidth modulators can be easily combined to fully exploit the performance of lasers~\cite{2016+196+213}. As a result, oscillation frequencies and repetition rates are controlled by using such actuator elements or by regulating the cavity length by moving end mirrors or cutting/fusion-splicing optical fibers. In contrast, optical microresonators are usually made of a single dielectric material so that their waveguides can strongly confine light in a tiny mode space~\cite{Kippenberg2011microresonator}. In return, this ultimately elemental configuration makes it a challenge to freely tune the OFC properties.
	
 Mechanical~\cite{PhysRevX.3.031003,Liu2020} and electrical tuning~\cite{Jung:14,Yao2018} can allow for dynamically control and stabilize the microcombs, and they can readily act as fast actuators for the comb frequencies in return for the limited spectral tuning bandwidth. Similarly, resonance control via thermal effects has also been implemented with an auxiliary laser coupled to a different resonance mode~\cite{Jost:15,Zhang2019}, pump power control using an intensity modulator~\cite{Yi:16,PhysRevLett.122.013902,Lucas2020}, or a Peltier element installed in a cavity-packaged module~\cite{Suh:19}. Although these techniques are suitable for narrow range of resonance tuning and compensation for environmental perturbation, i.e., soliton initiation~\cite{Yi:16,Zhang2019} or long-term stabilization~\cite{PhysRevLett.122.013902,Suh:19,Lucas2020}, these studies have paid little attention to the capability of the spectral tuning bandwidth. In this regard, temperature control of the entire resonator system using integrated microheaters has been adopted as the most effective and useful method of frequency tuning and resonance control for on-chip resonators operating mainly with a comb mode spacing exceeding 200~GHz~\cite{Miller:15,xue2015mode,Joshi:16,Xue:16,Kuse:20,Okawachi:22}. Despite the rapid progress made on larger mode spacing combs, broadband tunability and flexibility of soliton microcomb sources with a repetition rate in the microwave X- and K-bands (8-27~GHz) have not been demonstrated. These frequency bands are important  because they constitute the mainstream method for ultralow-noise photonic microwave generation with a direct optical-to-electric conversion process, and the microwave synthesis of the lowest-level phase noise signals has been achieved in millimeter-scale whispering-gallery-mode (WGM) microresonators~\cite{Lucas2020,Yang2021,Liang2015:high,Kwon2022}. Thus, widely tunable microcombs yielding the microwaves at these bands require a better understanding of the dynamics of soliton combs generated in WGM microresonators under thermal activation over a wide dynamic range.

    Here, we demonstrate and extend the spectral flexibility of a soliton microcomb with a 23~GHz repetition rate by exploiting temperature control and a detuning stabilization system. The center frequency, repetition frequency, and offset frequency of a soliton comb with a radio-frequency (rf) repetition rate are simultaneously tuned by thermal effects and stabilized by a feedback loop. We further show that the output power of a soliton comb is distinctly enhanced via thermally-induced coupling variation owing to the thermal expansion of a resonator coupled to a tapered fiber that ensures coupling flexibility. A qualitative analysis allows us to explain this interesting result. Moreover, we address the complex dynamics underlying soliton tuning arising from birefringence of crystalline magnesium fluoride ($\mathrm{MgF_2}$) resonators that gives rise to the difference in the thermal sensitivity of optical resonances belonging to different polarization modes. This feature leads to polarization-dependent tunability and has a unique effect on the soliton self-frequency shift (SSFS) via mode-coupling-induced dispersive wave (DW) emission.
	
	\section*{Results and Discussion}
	\subsection*{System overview}
The frequency tuning of cavity resonances in optical microresonators can be driven by two thermal effects, namely the thermo-optic and thermal expansion effects. The former originates from the change in refractive index and the latter corresponds to the volume change of the material in response to temperature variation, thus altering the optical length of the resonator. In general, both effects alter the resonance frequencies following a simple relation, $df/dT=-(\alpha_n+\alpha_l)f$, where $\alpha_n=(1/n)(dn/dT)$ and $\alpha_l=(1/l)(dl/dT)$ are the thermo-refractive and thermal expansion coefficients, respectively. The thermal coefficients in most resonator materials including crystalline $\mathrm{MgF_2}$ are positive, whereas some of the fluoride crystals (e.g., calcium fluoride, barium fluoride) or polydimethylsiloxane (PDMS) possess negative thermo-refractive coefficients. Although recent studies have demonstrated a potential advantage of the negative thermo-optic effect for microcomb generation~\cite{Lobanov:21,PhysRevA.103.023515}, it often results in resonance instability induced by thermo-optical oscillations~\cite{Diallo:15}. As a result, a crystalline $\mathrm{MgF_2}$ resonator is widely used to obtain a stable soliton state in thermal equilibrium~\cite{herr2014temporal,Liang2015:high,Lucas2018, Lucas2020, Fujii:22}. 

For uniaxial crystals such as $\mathrm{MgF_2}$, optical anisotropy should be considered~\cite{ghosh1998handbook,Fujii:20}. The use of a resonator made from $z$-cut $\mathrm{MgF_2}$ crystal results in slightly different refractive indices for the transverse-electric (TE) mode (extra-ordinary polarized, $n_e=1.382$) and the transverse-magnetic (TM) mode (ordinary polarized, $n_o=1.371$), and similarly the thermal coefficients exhibit anisotropy. In this work, we adopted the following values: $\alpha_{n,e}=0.23\times10^{-6}~\mathrm{K^{-1}}$, $\alpha_{n,o}=0.64\times10^{-6}~\mathrm{K^{-1}}$, and $\alpha_{l}=8.9\times10^{-6}~\mathrm{K^{-1}}$ at room temperature~\cite{ghosh1998handbook}. We also note that the thermal expansion can mainly be attributed to the ordinary direction $\alpha_{l,o}$, which corresponds to the expansion of the radial direction of the resonator~\cite{Lim2019}. Strictly speaking, thermal expansion coefficients are given as a function of temperature, but the impact is considered negligibly small in this study.

\begin{figure*}
	\centering
		\includegraphics[width=\textwidth]{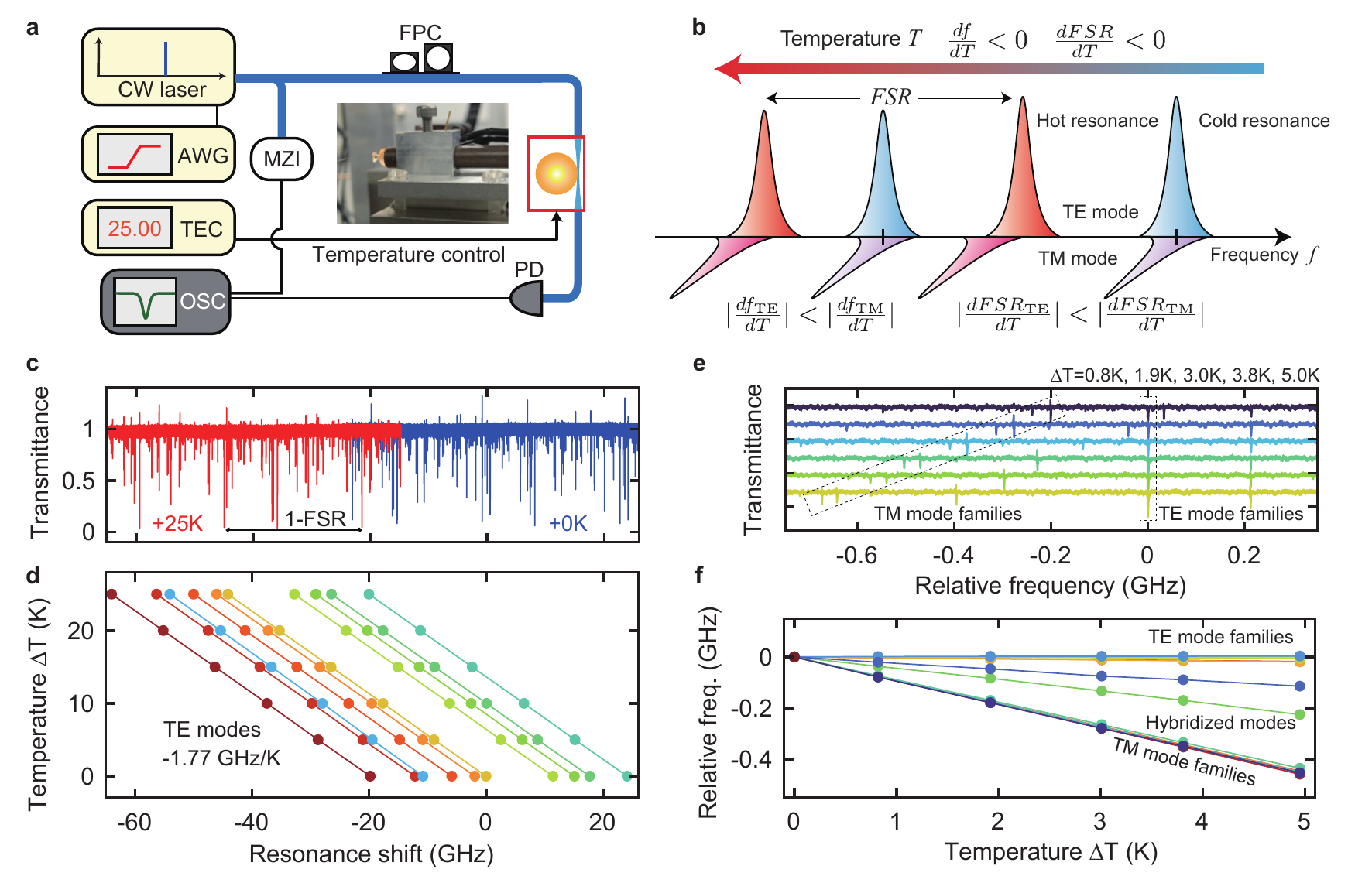}
		\caption{\label{fig1} Resonant frequency tuning efficiency for orthogonally polarized modes in a crystalline $\mathrm{MgF_2}$  microresonator. (a) Schematic experimental setup. Transmission spectra are recorded using a photodetector (PD) and an oscilloscope (OSC). The polarization of a continuous-wave (CW) laser is controlled by using a fiber polarization controller (FPC). The frequency axis of the transmission spectrum is calibrated with the reference signal of a Mach-Zehnder interferometer (MZI). AWG: Arbitrary waveform generator, TEC: Thermoelectric cooler. (b) Orthogonally polarized modes exhibit a different tuning efficiency with a temperature increase. Mode families belonging to the transverse-magnetic (TM) modes show greater efficiency than transverse-electric (TE) modes due to the difference in the thermo-optic coefficient. (c) Transmission spectra at $\Delta T=$0~K (blue) and $\Delta T=$25~K (red). (d) Extracted resonance shifts for 10 different modes in TE-polarization, which yields an average efficiency of $-1.77$~GHz/K. (e) The relative frequency shifts of TM mode families to the TE mode when the temperature is increasing by 5.0~K. (f) Extracted relative resonance shifts to one of the TE modes.}
	\end{figure*}

Figures~\ref{fig1}(a) and \ref{fig1}(b) show the experimental setup and concept of thermal resonance tuning in an $\mathrm{MgF_2}$ microresonator. While the tuning efficiency of resonance frequencies is independent of the resonator diameter, that of the mode-spacing, i.e., the free-spectral range (FSR), depends on the native mode spacing: $df_\mathrm{FSR}/dT = -c(\alpha_n+\alpha_l)/(2\pi n_g R)$, where $c$ is the speed of light in a vacuum, $n_g$ is the group index, and $R$ is the effective radius of a resonator. Thus, the tuning efficiency of the FSR corresponds $-213.2$~kHz/K for the TE mode and $-222.8$~kHz/K for the TM mode for the 23.35~GHz-FSR resonator that we used for soliton generation as described in the following section. The tuning efficiency of the resonant frequencies corresponds to $-1.765$~GHz/K for the TE mode and $-1.845$~GHz/K for the TM mode in the 1550~nm band. Figure~\ref{fig1}(c) and \ref{fig1}(d) show the transmission spectra and resonance shifts with the resonator temperature increasing from 300~K ($\Delta T=$~0~K) to 325~K ($\Delta T=$~25~K), respectively. The fitted slopes yield an average efficiency of $-1.77$~GHz/K, which indicates that these modes belong to the TE mode families. We also clearly observe the polarization dependence of the resonance tuning efficiency as shown in Figs.~\ref{fig1}(e) and \ref{fig1}(f). By offsetting the frequency shift of one of the TE modes, we can recognize the mode families exhibiting different tuning efficiencies as the TM modes, and the difference is $\Delta f_{\mathrm{TM}-\mathrm{TE}}\approx80$~MHz/K. It is sometimes difficult to assign the polarization for WGM resonators; however, a simple measurement of the amount of thermal shift readily distinguishes between the polarizations in $\mathrm{MgF_2}$ resonators. Interestingly, a few measured resonances show intermediate efficiency between the TE and TM modes due to the degeneracy of the effective indices, which forms a hybridized mode between orthogonal polarization modes~\cite{Ramelow:14}.

\subsection*{Versatile tuning of soliton microcombs}
\begin{figure*}
	\centering
		\includegraphics[width=\textwidth]{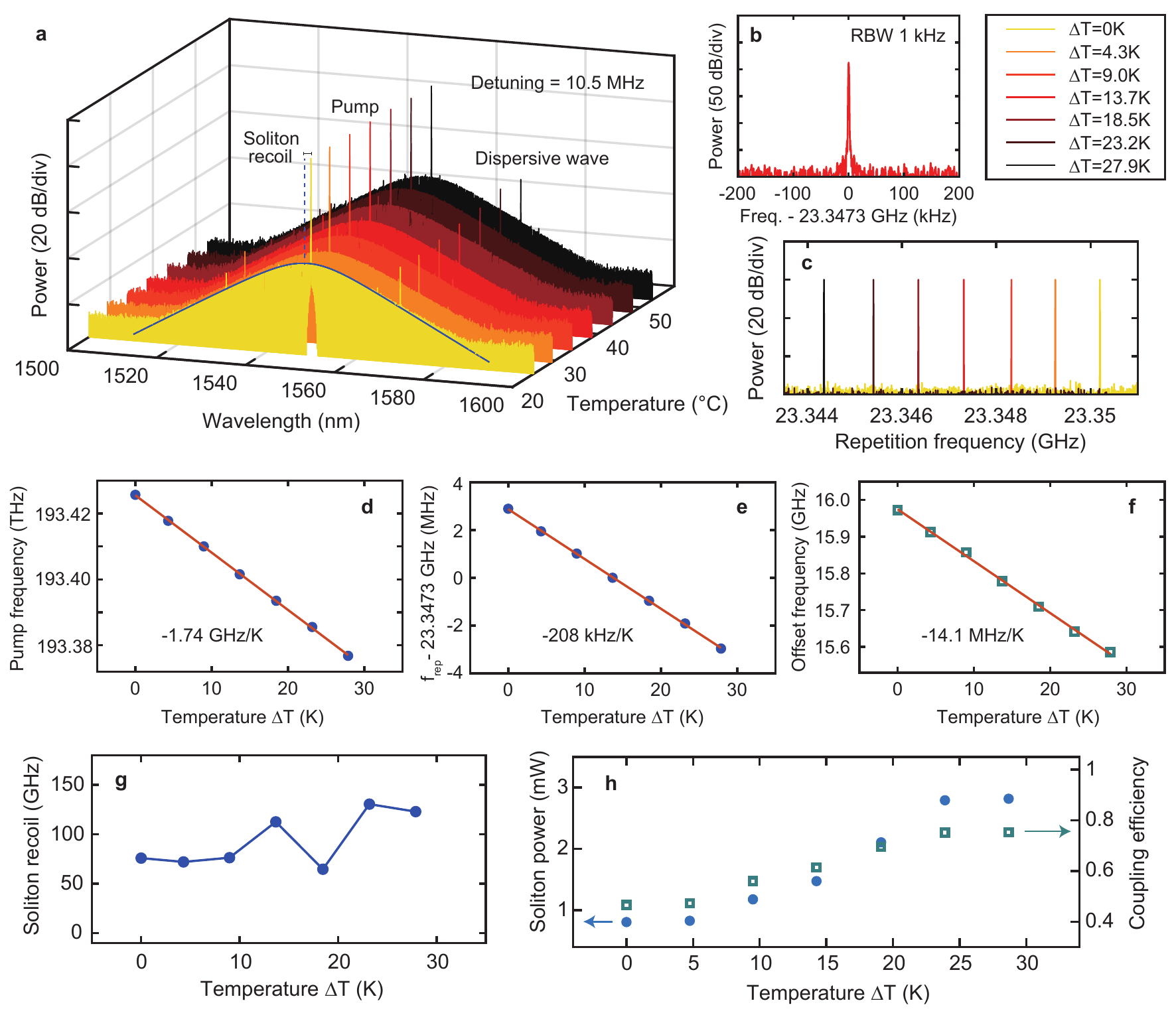}
		\caption{\label{fig2} Active thermal tuning of a soliton frequency comb. (a) A sequence of optical spectra of a single soliton state as a function of temperature with a fixed detuning of 10.5~MHz. The fitted envelope (blue line) indicates that the spectral center is shifted to the pump frequency (soliton recoil). (b, c) Radio-frequency beat-note spectrum yielding a soliton repetition frequency at 311.8~K ($\Delta T=$13.7~K) and temperature-dependent beat signals (d-f) Results of soliton tuning for the pump frequency, repetition frequency, and extracted carrier-envelope offset frequency, which yield the efficiency of each comb property. The solid red lines correspond to the results of linear regression. (g) Dependence of the soliton recoil frequency on temperature change. (h) Measured average soliton power (blue dots) and calculated coupling efficiency $\eta$ (green symbols) versus temperature change. }
	\end{figure*}

\begin{figure}
	\includegraphics{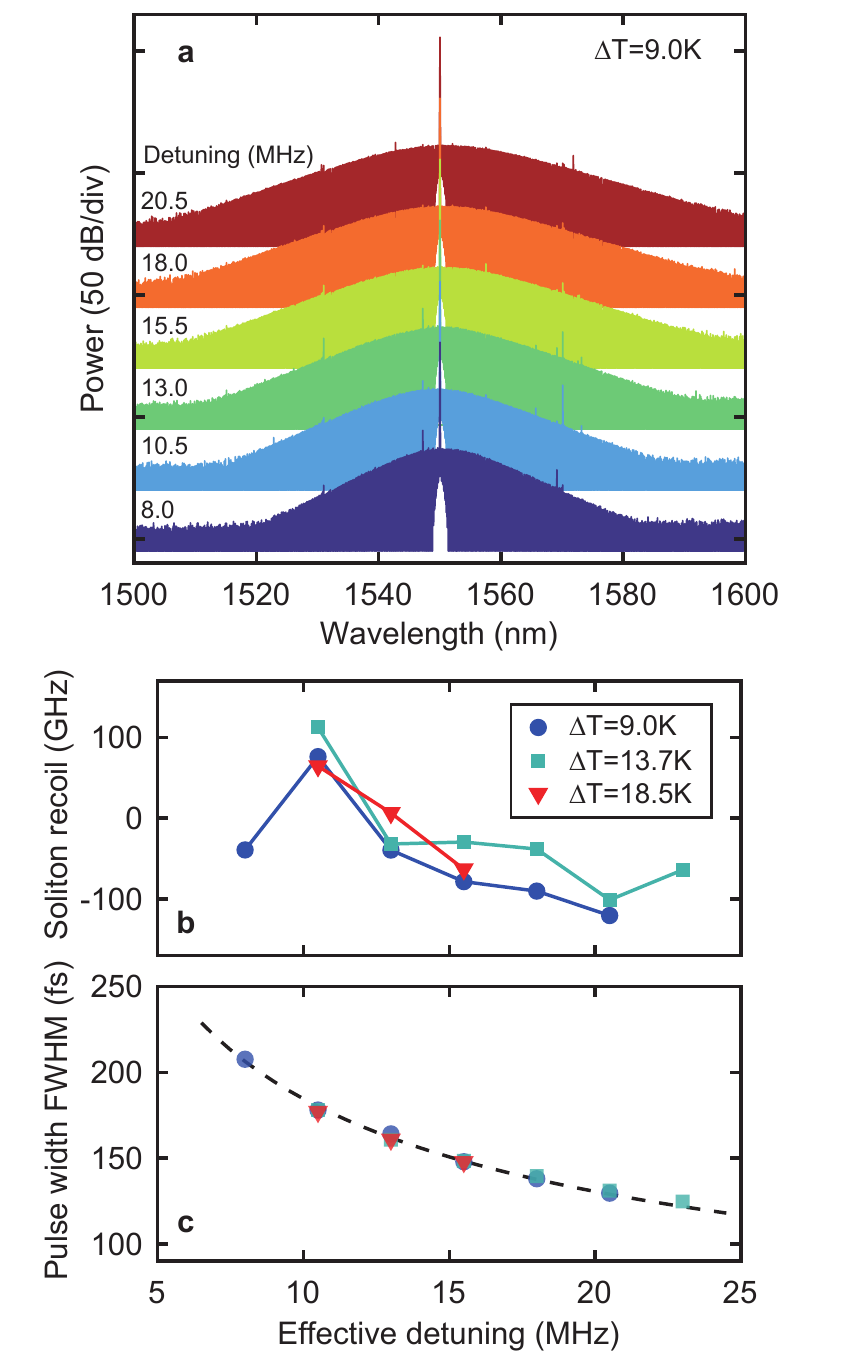}
	\caption{
		\label{fig3} Observed soliton properties with different effective detuning and temperature. (a) Optical spectra of a single soliton state with different detuning at 34~$^\circ$C ($\Delta T=9.0$~K). The effective detuning is scanned from 8~MHz (blue) to 20.5~MHz (red) in 2.5~MHz steps. (b,c) Measured recoil frequency and soliton pulse width derived from a $\mathrm{sech^2}$-fit versus the effective detuning. The positive (negative) value of the recoil frequency corresponds to the blue (red) shift with regard to the pump frequency. A dashed black line for the pulse width represents a theoretical curve. }
	\end{figure}

We perform the spectral tuning of a soliton microcomb using a temperature-controlled $\mathrm{MgF_2}$ microresonator with an FSR of ~23.35~GHz. A thermoelectric cooler (TEC) element is used to control the resonator temperature with 0.01~K accuracy with the appropriate feedback control, which enables active thermal tuning. First, we generate a soliton comb at 25~$^\circ$C ($\Delta T=0$~K) and measure the soliton beat note, which directly yields the soliton repetition rate. Here, the pump detuning is flexibly controlled by applying a modulation frequency to a phase modulator. We obtain temperature-dependent optical and electrical spectra as shown in Fig.~\ref{fig2}, while pump detuning dependence is presented in Fig.~\ref{fig3} (see Methods and Supplementary Note~1 for the details of our experimental setup). The pump detuning is an essential parameter for determining both the fundamental properties of a soliton comb such as the soliton pulse width and the soliton noise limit, which involves the soliton spectral recoil induced by DWs and the Raman effect~\cite{Yi2017,Lucas2017,Yang2021}. 

Figure~\ref{fig2}(a) shows the measured optical spectra of single-soliton states when the temperature is increased up to 53~$^\circ$C with a fixed detuning of 10.5~MHz. The frequencies of each comb line are defined by the well-known relation, $f_{m}=f_{\mathrm{ceo}} +m \times f_{\mathrm{rep}}$, where $f_{m}$ is the frequency of each comb mode and the index $m$ is an integer. Since the repetition rate $f_{\mathrm{rep}}$ is directly obtained as the beat frequency as shown in Figs.~\ref{fig2}(b) and \ref{fig2}(c), it is easily possible to extract the carrier-envelope offset frequency $f_{\mathrm{ceo}}$ by subtracting the pump frequency from an integer multiple of $f_{\mathrm{rep}}$  ($m=6365$). Figures~\ref{fig2}(d)-\ref{fig2}(f) show the variation in the pump (center) frequency $f_p$, $f_{\mathrm{rep}}$, and  $f_{\mathrm{ceo}}$, where the fitting slope (solid line) yields the tuning efficiency, $-1.74$~GHz/K, $-208$~kHz/K, and $-14.1$~MHz/K, respectively. The total tuning ranges of $f_p$, $f_{\mathrm{rep}}$, and $f_{\mathrm{ceo}}$ reach $-48.8$~GHz, $-5.85$~MHz, and $-386$~MHz, respectively, with 28~K temperature change.  The measured tuning efficiencies agree well with the theoretical predictions discussed in the previous section, and the results confirm that the pump mode belongs to the TE modes. It should also be noted that the total tuning range of the pump frequency exceeds 2-FSRs owing to the narrow mode spacing of our millimeter-scale resonator. This feature makes it possible to search for a desired resonance mode by using a thermal heating method instead of broadband laser frequency control, whose tuning range depends on the performance of the pump laser and may be insufficient to access all resonance modes within 1-FSR. 

To investigate the effect of thermal tuning on soliton properties in detail, we then plot the recoil frequency and average soliton power as a function of temperature (Fig.~\ref{fig2}(g) and \ref{fig2}(h)). We draw attention to the fact that temperature change leads to a variation in the soliton recoil frequencies even with the same detuning values (Fig.~\ref{fig2}(g) and Fig.~\ref{fig3}(b)). Since the Raman effect can be neglected in $\mathrm{MgF_2}$ crystals because of the very narrow gain bandwidth, DW emission, which results from higher-order dispersion and mode couplings, must critically affect the degree of soliton recoil~\cite{Yang2021,Lucas2017,Okawachi:22}. Even though the spectral positions of the DWs are inherent features in a single microresonator with fixed detuning, we find that the intensity and positions of strong peaks vary irregularly with temperature, and this indicates that changes in resonator temperature have a strong influence on resonator mode structure, i.e., thermally-induced resonance shift. Specifically, birefringent properties, which affect the tuning efficiency for orthogonal polarization, induce considerable mode hybridization rather than linear frequency shifts of all the resonant modes via strong cross-polarization mode coupling~\cite{Ramelow:14}.  The design and prediction of the spectral position of the DWs are still challenges~\cite{Bao2017}; nevertheless, we believe that thermal activation could be utilized for the active tuning of the DWs in a WGM resonator supporting many different transverse modes.

More surprisingly, the soliton power is significantly increased with the temperature change even though a tapered fiber is in contact with the resonator and the position is not adjusted during the series of measurements. This result is not intuitively anticipated because the changes in the pump frequency and repetition frequency have little impact on the average soliton power given by $P_{\mathrm{sol}}=\frac{2\eta A_\mathrm{eff}}{n_2Q}\sqrt{-2n c\beta_2 \delta\omega}$~\cite{yi2015soliton,Lucas2017,Fujii:22}, where $\eta$ is the coupling efficiency, $A_{\mathrm{eff}}$ is the effective mode area, $n$ is the refractive index, $n_2$ is the nonlinear index, and $Q$ is the total decay rate.
Although the pump detuning $\delta\omega=(\omega_0-\omega_p)$ and resonator group-velocity dispersion $\beta_2$ contribute to the soliton power, we can decouple them because the measured pulse widths $\tau = \sqrt{-{c\beta_2}/{2n\delta\omega}}$ are almost constant ($\tau \approx178$~fs) under a fixed detuning of 10.5~MHz, which is direct evidence that the overall dispersion is unchanged.  In fact, the pulse widths follow the theoretical prediction even for different temperatures at the same detuning value (Fig.~\ref{fig3}(c)). 
When we neglect the changes in the mode area, refractive index and nonlinear refractive index for a temperature increase of 28~K, another possibility is that the significant change in the soliton power is caused by the variation in the coupling rate $\eta/Q~(={n_2 P_{\mathrm{sol}}}/{2 A_\mathrm{eff} \sqrt{-2n c\beta_2 \delta\omega}})$.
The green symbol in Fig.~\ref{fig2}(h) represents the extracted coupling efficiency $\eta=\kappa_{\mathrm{ext}}/(\kappa_{\mathrm{int}}+\kappa_{\mathrm{ext}}$) as a function of temperature change at a constant intrinsic cavity decay rate ($\kappa_{\mathrm{int}}/2\pi=0.43$~MHz). This result indicates that the coupling strength, namely $\kappa_{\mathrm{ext}}/2\pi$, is eventually increased 3.4~times (0.38~MHz to 1.3~MHz) by heating the resonator despite the fact that neither the position nor the diameter of the fiber, which could alter the coupling strength, is adjusted during the measurement. It should be noted that the accessible soliton step varies due to the change in the fiber-to-resonator coupling condition as the cavity temperature increases. We emphasize that the coupling strength is known to be determined by the mode overlap integral and phase mismatch between a resonant mode and a fiber mode. In this respect, we theoretically detail the mechanism of the thermally-induced coupling variation.

\subsection*{Analysis of dynamic change in coupling strength}
		
\begin{figure*}
	\centering
		\includegraphics[width=\textwidth]{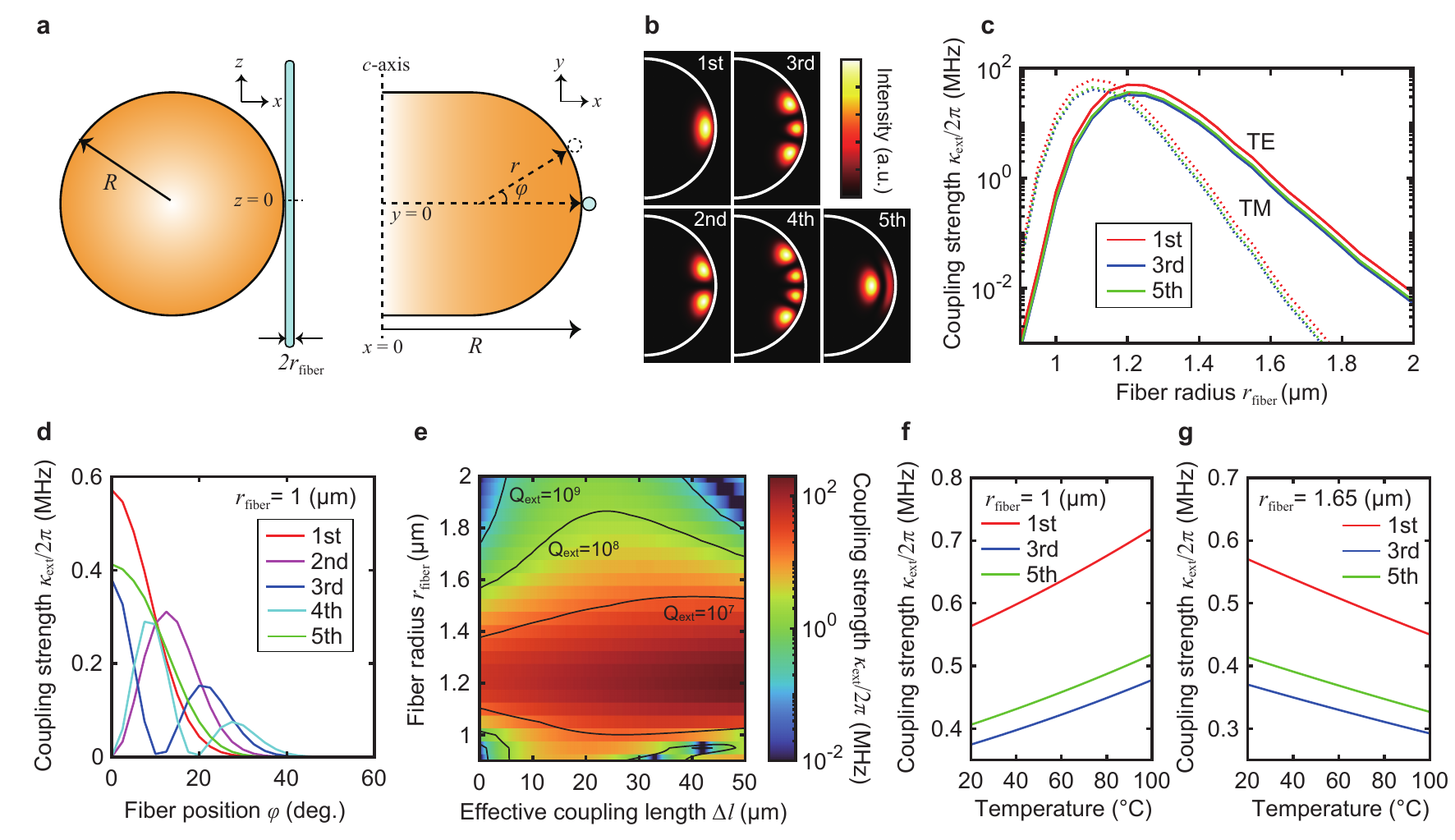}
		\caption{\label{fig4} Analysis of the coupling strength in a crystalline $\mathrm{MgF_2}$ microresonator coupled to a silica tapered fiber. (a) Schematics of the fiber-resonator coupling model in $x$-$y$-$z$ coordinates, where $R$ and $r$ correspond to a resonator radius and curvature, respectively. The fiber position along the curvature is given by the polar angle $\varphi$. (b) Mode profiles of several whispering-gallery (WG) modes. The fundamental mode is designated as 1st and the higher-order modes are named sequentially. The color bar shows the electric field intensity. (c) Simulated coupling strength versus fiber radius for three WG modes. The fiber radius showing the maximum coupling strength modes is shifted literally due to the difference between refractive indices of the transverse-electric (TE) and transverse-magnetic (TM) modes. (d) Coupling strength as a function of fiber position $\varphi$ for the TE modes. (e) Contour map of coupling strength showing the dependence on fiber radius and effective coupling length $\Delta l$ for the TE 1st mode. The color bar shows the coupling strength $\kappa_{\mathrm{ext}}/2\pi$ in units of MHz. (f, g) Coupling strength as a function of temperature for the 1st, 3rd, and 5th TE modes at different fiber radii, (f) for $r_{\mathrm{taper}}=$1~\textmu m and (g) for $r_{\mathrm{taper}}=$1.65~\textmu m. The temperature-dependent refractive index change modifies the phase-matching condition, resulting in increased or reduced coupling strength. }
	\end{figure*}

In light of the experimental results, we numerically analyze the effect of the thermal activation on the coupling strength between a WGM resonator and a silica tapered fiber coupler (The calculations are detailed in Methods and Supplementary Note~2-4). Figure~\ref{fig4}(a) shows a schematic model of our simulation for the coupling strength in a millimeter-scale $\mathrm{MgF_2}$ resonator. On the assumption of a weak coupling approximation (i.e., the coupling coefficient is sufficiently small compared with the propagation constant), the coupling coefficient can be expressed as~\cite{HUMPHREY2007124},
	\begin{equation}
	\begin{split}
		\tilde{\kappa}=&\frac{\omega \epsilon_0}{4}(n_\mathrm{r}^2-n_0^2) N_f N_r \\ &\times \iiint \left( \bm{E}_\mathrm{f} \cdot \bm{E}_\mathrm{r}^{*}\right) e^{i \Delta \beta z}   dx dy dz \label{kappa}
	\end{split}
	\end{equation}
where $\bm{E}_\mathrm{f(r)}$ corresponds to the complex electric field amplitudes of the fiber (resonator), $N_\mathrm{f(r)}$ is the normalization constant for the fiber (resonator), and $n_\mathrm{r}$ and $n_0$ are the refractive indices of the resonator and air. The difference between the propagation constants $\Delta \beta = \beta_\mathrm{f}-\beta_\mathrm{r}$ contributes to the phase mismatch between the fiber and resonator modes. The integration is performed over a volume of the resonator in a three-dimensional Cartesian coordinate system, where the $z$-axis represents the length direction of a tapered fiber. The coupling strength is then obtained from the relation $\kappa_\mathrm{ext} \approx |\tilde{\kappa}|^2(c/2\pi n_\mathrm{r} R)$, which accordingly corresponds to $\omega/Q_\mathrm{ext}$. 
 Importantly, Eq.~(\ref{kappa}) indicates that both the overlap integral between the fiber and resonator modes and the phase-matching condition are central to the coupling strength. As a hypothesis, the thermal expansion effect can modify the coupling strength as a result of the variation in the mode overlap, which is determined by the degree of contact between the fiber and resonator, and the thermo-optic effect changes the effective refractive indices, thus altering the phase-matching condition.

To investigate the dynamics of coupling variation, we first begin with numerical simulation of the coupling strength with respect to fiber diameter for different transverse modes when the fiber and resonator are placed in contact at one point. Figures~\ref{fig4}(b) and \ref{fig4}(c) present the mode field distribution for featured five transverse modes and the calculation results. As mentioned earlier, $\mathrm{MgF_2}$ resonators exhibit different refractive indices for orthogonal polarized modes, and therefore the fiber diameter which minimizes the phase mismatch also differs for the TE and TM mode families. In addition, higher-order polar and radial modes exhibit different coupling strengths due to the difference of the propagation constant and mode profiles. The two modes (2nd and 4th) of the five WG modes are not shown in Fig.~\ref{fig4}(c) because the coupling strengths are several orders of magnitude smaller, and coupling does not occur. This is because the fiber is positioned at $y=0$ in this simulation, thus significantly reducing the mode overlap.  The coupling strength for varying the fiber position, defined as polar angle $\varphi$, is presented in Fig.~\ref{fig4}(d), where the fiber position is moved along a resonator curvature. This result confirms that higher-order WG modes can also be efficiently excited by changing the contact position of the fiber. Indeed, the coupling strength can be controlled by carefully adjusting the contact position, fiber thickness, and effective coupling length, namely the degree of contact with a resonator, in practical experiments.

To consider the effect of the effective coupling length $\Delta l$ instead of single point coupling at $z=0$, we modified Eq.~(\ref{kappa}) by including an additional coupling term $\tilde{\kappa}_{\Delta l}$ supposing that the mode overlap is maximized at $z=0$ and remains constant in the finite coupling regime $\Delta l$. Then, the additional coupling term is given by $\tilde{\kappa}_{\Delta l} = \tilde{\kappa}_{z=0} \sin({\Delta \beta \Delta l/2})/(\Delta \beta/2)$~\cite{Hosseini:10}. Such an approach is referred to as a pulley-style configuration and widely employed with integrated resonators to precisely engineer the coupling strength~\cite{Hosseini:10,PhysRevApplied.7.024026,Lu2019}. The significant difference is that the tapered fiber used in our system provides a considerable degree of coupling flexibility in contrast to integrated resonators with a fixed waveguide as mentioned earlier. The dependence of the coupling strength on effective coupling length and fiber diameter is shown in Fig.~\ref{fig4}(e) as a contour map. The coupling strength exhibits a periodic dependence on the effective coupling length, which originates from a relative phase shift between the fiber and resonator modes as seen in a directional coupler~\cite{doi:10.1063/1.106188}. When the fiber diameter is 3.3~\textmu m (typical diameter of a tapered fiber is 2--4~\textmu m in our experiments), the coupling strength for the fundamental (1st) TE mode is gradually increased from $\kappa_{\mathrm{ext}}/2\pi\approx$ 0.56 to 2.7~MHz with $\sim$10~\textmu m coupling length change. The results in the fundamental TM mode are also presented in Supplementary Note~5.

Our interest here is whether or not the thermal effects indeed modify the coupling strength. Since the temperature-dependent resonator expansion is given by $\Delta R(\Delta T) = \alpha_l R \Delta T$, the radius of the resonator ($R=1400$~\textmu m) is increased by $\sim$0.5~\textmu m with 30~K heating. It should be noted that the change in the effective coupling length is a difficult parameter to assess experimentally because it relies on the tensile strength of the tapered fiber and the actual tapered length. Nevertheless, we emphasize that  the effective coupling length is likely to increase by several microns or several tens of microns in practical experiments (See Supplementary Note~3), and therefore, the thermal expansion effect (i.e., the change in the effective coupling length) could cause a dynamic, wide range coupling enhancement as seen in our experimental results in Fig.~\ref{fig2}. 
We further mention that the thermo-optic effect also contributes to the coupling strength via the phase-mismatch condition $\Delta \beta$ due to the considerable difference in the thermo-optic coefficients of silica and $\mathrm{MgF_2}$ crystal. In thermal equilibrium, the temperature of a silica tapered fiber is considered to be the comparable to that of the resonator in the region where the fiber is in contact with the resonator. However, we find that the phase-mismatch induced by the thermo-optic effect only alters the coupling strength of up to $\sim$20\% with a 50~K temperature change (Fig.~\ref{fig4}(f,g)) and is thus less effective than the expansion effect.
	
\section*{Conclusions}
In summary, we have experimentally demonstrated the versatile tuning of rf repetition-rate soliton combs in a crystalline $\mathrm{MgF_2}$ microresonator. The main properties of frequency combs, namely, the center frequency, repetition frequency, and carrier-envelope offset frequency are tuned by $-48.8$~GHz, $-5.85$~MHz, and $-386$~MHz, respectively, by using the temperature control of a fiber-coupled resonator system. Additionally, we show the possibility of spectrally tuning the DWs accompanied by soliton recoil via the dynamic modification of cavity mode structures due to the birefringence of crystalline microresonators.
Thermally-induced coupling variation that leads to pronounced soliton power enhancement has also been demonstrated and studied theoretically to validate this effect. So far, optimization of a coupling position and a fiber thickness has been performed  based on practical experiences and patient adjustment rather than a theoretical method. The quantitative analyses also provide helpful guideline for the rational optimization of this “rule of thumb” fiber-resonator coupling to multi-mode WGM resonators. 

Previous studies have not reported such a significant coupling change induced by thermal effects~\cite{Miller:15,Xue:16,Kuse:20,Fujii:22,Okawachi:22}. In our case, the flexibility of the tapered fiber coupler and the relatively large thermal expansion coefficient of the crystalline material (e.g. $\mathrm{MgF_2}:8.9\times10^{-6}~\mathrm{K^{-1}}$. fused $\mathrm{SiO_2}:0.6\times10^{-6}~\mathrm{K^{-1}}$, $\mathrm{Si}:2.6\times10^{-6}~\mathrm{K^{-1}}$~\cite{ghosh1998handbook}) allow modification of the coupling strength, and resulting in the enhancement of the soliton comb power. Although the implementation of thermally-induced dynamic coupling control is a challenge, it would be possible if the resonator and the waveguide have very different thermo-optic coefficients. For future practical applications, broad and reliable soliton tuning could be employed in microcomb systems. In particular, photonic rf oscillators exploit the intrinsic low phase-noise properties of soliton combs in ultrahigh-Q crystalline resonators~\cite{Lucas2020}, whereas the thermal drift of optical resonances leads to the degradation of the phase-noise performance in the low Fourier frequency regime. Thus, temperature control and subsequent high-level stabilization~\cite{Savchenkov:07,Lim2017} are essential for mitigating the thermal drift and pursuing the fundamental noise limit of soliton microcombs~\cite{Drake2020}. This study provides a greater understanding regarding wide-range tunability of the comb frequency and output power towards the implementation of both a high-purity photonic oscillator and a high-coherence laser source for optical communications and dual-comb spectroscopy that requires precise control of the difference between repetition frequencies.

	\section*{Methods}
	\subsection*{Experimental setup}
An $\mathrm{MgF_2}$ microresonator is fabricated from a $z$-cut crystalline disk by shaping and hand polishing with diamond slurry. The FSR of 23.35~GHz corresponds to a major radius of 1.49~mm and typical loaded Q-factors are $10^8$--$10^9$ in the 1550~nm band. The curvature radius is approximately $50$~\textmu m. The group-velocity dispersion of the pumped mode in our work is $\beta_2=d^2\beta/d\omega^2\approx-6.3~\mathrm{ps^2/km}$, yielding anomalous dispersion.  The group-velocity dispersion is defined as the derivative of the inverse group velocity with respect to the angular frequency~\cite{FujiiTanabe+2020+1087+1104}. The crystalline resonator is fixed with adhesive to a metal rod a few centimeters long, so that the temperature of the resonator is controlled through the metal parts with a temperature transducer. For soliton generation, a continuous-wave fiber laser (linewidth$<$0.1~kHz) is amplified by an erbium-doped fiber amplifier (EDFA), and an optical power of $\sim$400~mW is coupled via a silica tapered fiber. The length of the tapered region is $\sim$40~mm, and the diameter of the tapered fiber and the contact position with the resonator is optimized to make it possible to observe distinct soliton steps. Typically, we employed fiber diameters of 2--4~\textmu m, and brought the fiber into contact with the resonator to realize stable light coupling. A single soliton is generated via a forward laser frequency sweep with an adequate scan speed (0.1--10~MHz/ms), and then the soliton is captured by activating a servo controller to lock the effective detuning via a phase-modulation technique~\cite{PhysRevLett.122.013902,PhysRevLett.121.063902,Fujii:22}. The feedback control of the pump laser frequency enables us to stabilize the soliton state against the thermal drift of the cavity resonance and the long-term frequency fluctuation of the pump laser~\cite{PhysRevLett.121.063902}. The absolute frequency of the pump laser is monitored by using a wavelength meter, whose spectral resolution (0.1 pm) limits the absolute accuracy of $f_p$ and $f_\mathrm{ceo}$.

\subsection*{Numerical analysis of resonator and fiber modes}
Mode profiles of the WGM resonator and tapered fiber are calculated by conducting a finite element simulation and a theoretical analysis, respectively. We perform a finite element method (FEM) simulation by using FEM software (COMSOL Multiphysics) to calculate the resonator modes. For the simulation, we model an axisymmetric WGM resonator with a resonator radius of 1400~\textmu m and a curvature of 25~\textmu m, respectively, with a calculation area of 55~\textmu m by 55~\textmu m. The refractive indices of $\mathrm{MgF_2}$ for the TE and TM modes are set independently for the simulation.  

To calculate the mode overlap integral between the fiber and resonator modes, the three-dimensional electrical components are mapped on Cartesian-coordinate grids and then normalized by a factor, given as,
	\begin{equation}
	    1=\frac{1}{2}N_\mathrm{f(r)}^2 \sqrt{\frac{\epsilon_0}{\mu_0}} \iint n_\mathrm{f(r)}(x,y) |\bm{E}|_\mathrm{f(r)}^2 dx dy.
	\end{equation}
The $y$- and $x$-axes correspond to the dominant directions of the electrical field of the TE, and TM modes, respectively. The mode field of a tapered fiber can be analytically obtained by solving the wave equation in the cylindrical coordinates, where the propagation direction is taken to be the $z$-direction and the center of the contact position is defined as $z=0$. A detailed derivation is presented in Supplementary Note~2. The equation gives several fiber modes with different propagation constants, whereas we only focus on the fundamental fiber mode, i.e., $\mathrm{HE}_{11}$ mode, which is linearly polarized single-maximum mode for the simulation. The refractive indices for a fiber and surrounding air are set at $n_\mathrm{f}=1.444$ and $n_0=1.000$ at 25$^\circ$C. 

	\subsection*{Calculation of coupling coefficient}
The coupling coefficient is obtained by performing a three-dimensional integration of Eq.~(\ref{kappa}). The integration for the $x$--$y$ plane, which is transverse to the propagation direction, is performed over the cross-sectional area of an $\mathrm{MgF_2}$ microresonator at each $z$-point. Because of the resonator curvature $r$, the overlap integral gradually decreases in accordance with the light propagation in the $z$-direction. Then, the overlap integral is further integrated along the $z$-axis to take the phase-matching condition into account. A sufficiently large integral interval is chosen so that the coupling coefficient converges to a constant value. When we calculate the additional coupling introduced by the effective coupling length, the section is considered to be a straight waveguide for simplicity, which can be justified because the resonator radius ($R=1400$~\textmu m) is much larger than the diameter of a tapered fiber ($\sim$2--4~\textmu m). The effective refractive index for a resonator mode is calculated from the electric field distribution and temperature-dependent refractive indices for extraordinary and ordinary rays (See Supplementary Note~3 for details).

\section*{Acknowledgments}
This work is supported in part by JSPS KAKENHI (JP19H00873, JP22K14625). S.F. acknowledges support from RIKEN Special Postdoctoral Researcher Program. The authors thank Dr.~W.~Yoshiki for supporting numerical simulation.

% \section*{Data availability}
% The data that support the plots of this paper and other findings within this study are available from the corresponding author upon reasonable request.

% \section*{Code availability}
% The codes used for this study are available from the corresponding author upon reasonable request.

\section*{Author contributions}
S.F and K.W. performed the experimental measurements. S.F. and R. S. conducted the numerical simulations. K. W. fabricated the crystalline microresonators, and S. F., K. W., H. K. and S. K. developed the experimental setups. S. F. and K. W. analyzed the data. S. F wrote the manuscript with input from Y. K. and T. T.. All authors had constructive discussions. S. F. and T. T. supervised the project. 

% \section*{Competing interests}
% The authors declare no competing interests.

\section*{References}
	\bibliography{soliton_tuning}% Produces the bibliography via BibTeX.
 
\end{document}